\documentclass[conference]{IEEEtran}
\newcommand\myVSpace[1][10pt]{\rule[\normalbaselineskip]{0pt}{#1}}
\IEEEoverridecommandlockouts
\usepackage{cite}
\usepackage{amsmath,amssymb,amsfonts}
\usepackage{algorithmic}
\usepackage{graphicx}
\usepackage{textcomp}
\usepackage{xcolor}
\usepackage{algorithm}
\usepackage{hyperref} 
\usepackage{algorithmic}
\def\BibTeX{{\rm B\kern-.05em{\sc i\kern-.025em b}\kern-.08em
    T\kern-.1667em\lower.7ex\hbox{E}\kern-.125emX}}
    
\begin{document}

\title{Avoiding convergence stagnation in a quantum circuit evolutionary  framework through an adaptive cost function}

\author{\IEEEauthorblockN{1\textsuperscript{st} Bruno O. Fernandez}
\IEEEauthorblockA{\textit{QuIIN - Quantum Industrial Innovation} \\
\textit{SENAI CIMATEC}\\
Salvador, Brazil \\
brunoozielf@gmail.com}
\and
\IEEEauthorblockN{2\textsuperscript{nd} Rodrigo Bloot}
\IEEEauthorblockA{\textit{ILACVN} \\
\textit{UNILA}\\
Foz do Iguaçu, Brazil \\
rgbloot@gmail.com}
\and
\IEEEauthorblockN{3\textsuperscript{rd} Marcelo A. Moret}
\IEEEauthorblockA{\textit{QuIIN - Quantum Industrial Innovation} \\
\textit{SENAI CIMATEC}\\
Salvador, Brazil \\
moret@fieb.org.br}
}

\maketitle

\begin{abstract}
Binary optimization problems are emerging as potential candidates for useful applications of quantum computing. Among quantum algorithms, the quantum approximate optimization algorithm (QAOA) is currently considered the most promising method to obtain a quantum advantage for such problems. The QAOA method uses a classical counterpart to perform optimization in a hybrid approach. In this paper, we show that the recently introduced method called quantum circuit evolutionary (QCE) also has potential for applications in binary optimization problems. This methodology is classical optimizer-free,  but, for some scenarios, may have convergence stagnation as a consequence of smooth circuit modifications at each generation. To avoid this drawback and accelerate the convergence capabilities of QCE, we introduce a framework using an adaptive cost function (ACF), which varies dynamically with the circuit evolution. This procedure accelerates the convergence of the method. Applying this new approach to instances of the set partitioning problem, we show that QCE-ACF achieves convergence performance identical to QAOA but with a shorter execution time. Finally, experiments in the presence of induced noise show that this framework is quite suitable for the noisy intermediate-scale quantum era.
\end{abstract}


\section{Introduction}
Problems that require some type of optimization occur regularly in science and engineering. Quantum algorithms of the variational type, also known as Variational Quantum Algorithms (VQA's), emerge as tools of possible use for solving such problems, which include study of molecules, general optimization, and machine learning (see, e.g., \cite{vqe_original,Kandala_2017,vqa,qsvm}). However, it is important to note that several critical problems in industry are modeled in terms of integer linear optimization problems, which can require a lot of computational resources when treated classically in large-scale applications.

Consequently, many binary optimization problems are NP-hard and candidates for quantum computer utility in the near future. With the advent of the noisy intermediate scale quantum (NISQ) era, terminology introduced in \cite{preskill2018quantum}, the search for algorithms well adapted to such devices has driven research in this area. The quantum approximate optimization algorithm (QAOA) introduced in \cite{faihi2014qaoa} is an algorithm adapted for NISQ that performs well on binary integer optimization problems such as MAX-CUT. Furthermore, with some variations, it has been tested in several scenarios with success (see, e.g., \cite{qaoa_image}\cite{RUAN202398}, \cite{QIAN2023}). 

A bottleneck in QAOA concerns its structure, which requires a classical optimization calculation, and resides in the convergence stagnation effect induced by a massive amount of local minima, as explained in \cite{cerezop} and \cite{Otfree}. However, QAOA's capabilities are still undergoing improvement. Recently, Cheng L. et al.\cite{cheng2024} managed to improve the optimization process of the parameters of QAOA using a classical gradient-free optimizer dubbed Double Adaptive-Region Bayesian Optimization.

Looking for a different approach to solve binary optimization problems, the variational methods free from classical optimization calculus seem to be a good alternative to avoid convergence stagnation phenomena. The Quantum Circuit Evolution method (QCE), introduced by Franken et al.\cite{QCEPROC} presents a variable circuit structure with arbitrary ansatz without the need to use classical optimizers. 

The procedure does not depend on Hamiltonian problem as QAOA and has variable depth. However, even without the use of classical optimizers, QCE may eventually experience convergence stagnation under certain circuit selections. With appropriate circuit selection and increasing the number of generations, the method can escape the local minimum. However, this leads to an increase in circuit complexity and a considerable increase in execution time, representing a disadvantage compared to QAOA.

In this paper, we avoid the difficulties of the QCE regarding convergence by introducing a framework that uses an adaptive cost function. The proposed approach significantly improves the convergence of the method with a considerable reduction in execution time. The technique was tested and compared with QAOA on instances of the set partitioning problem, demonstrating encouraging results.\\

\section{The target problem}

This section provides a concise introduction to the set partitioning problem, an NP-hard combinatorial optimization challenge frequently encountered across various practical applications. Typically, the problem involves partitioning a set of distinct elements into multiple subsets, where each element is exclusively assigned to exactly one subset. A prominent example of its application arises in airline operations, particularly in crew scheduling, which involves allocating pilots and attendants to crews in such a way that flight assignments meet all regulatory constraints, including qualifications and flight-time restrictions, while minimizing operational costs. Consequently, this problem serves as a critical tool in optimizing resource management for complex operational systems \cite{garey97}.

\subsection{Mathematical Formulation}

Consider a set of partitions denoted as \( P \), each partition \( p \in P \) associated with a non-negative cost \( w_p \), and a set of elements \( I \). The objective is to select partitions from the set \( P \) such that every element of the set \( I \) is covered exactly once, minimizing the total cost. To formulate this mathematically, a binary decision variable \( x_p \) is introduced for each partition \( p \), defined as 1 if the partition is chosen and 0 otherwise. The subset of elements contained within each partition \( p \) is represented by \( I_p \). Consequently, the mathematical model for this optimization problem is given by:

\begin{equation}
	\text{Minimize} \quad \sum_{p \in P} w_p x_p
\end{equation}

Subject to the constraints:

\begin{equation}
\sum_{p \in P:i \in I_p} x_p = 1, \quad \forall i \in I,
\end{equation}

\begin{equation}
x_p \in \{0,1\} \quad \forall p \in P
\end{equation}
The constraints above guarantee that each element is uniquely allocated to exactly one partition~\cite{VQE_SHORT}. For the application of quantum algorithms, the problem is typically transformed into a Quadratic Unconstrained Binary Optimization (QUBO) form \cite{Glover2022}.

\subsection{QUBO Formulation}

Applying penalty methods from optimization theory~\cite{OPT2019}, the constrained optimization problem can be reformulated into a QUBO model. This procedure involves adding penalty terms for constraint violations directly into the objective function. Consequently, the cost function given as a QUBO formulation is expressed as \cite{VQE_SHORT}:

\begin{equation}
	 \quad C(\mathbf{x})=\sum_{p \in P} w_p x_p + \sum_{i \in I} \rho_i\left(\sum_{p \in P; i \in I_p} x_p - 1
\right)^2,
\label{cost}
\end{equation}
where \( \rho_i \) are penalty parameters applied to each element \( i \in I \), ensuring that the original constraints are implicitly maintained. A proper selection of these penalties guarantees that the QUBO formulation remains equivalent to the original constrained formulation. This unconstrained formulation can subsequently be represented as an Ising Hamiltonian suitable for quantum computation \cite{andrew_lucas}.

\section {QAOA and QCE: A brief description}

In this section, we briefly introduce two quantum algorithms employed in combinatorial optimization: the Quantum Approximate Optimization Algorithm (QAOA) and the Quantum Circuit Evolution (QCE).

\subsection{Quantum Approximate Optimization Algorithm}

Introduced by Farhi et al.\cite{faihi2014qaoa}, QAOA employs parameterized quantum circuits whose objective is to approximate solutions by adjusting their parameters iteratively. The QAOA algorithm consists of two alternating unitary operations, parameterized by angles $\gamma$ and $\beta$. The depth  of the circuit affects the quality of approximation, where higher values of  generally yield better solutions at the cost of greater circuit complexity. A notable characteristic of QAOA is that its performance steadily improves as the parameter  increases, theoretically approaching the exact optimal solution for sufficiently large  values.

\subsection{Quantum Circuit Evolution}

The Quantum Circuit Evolution method, introduced by \cite{QCEPROC}, employs evolutionary strategies to optimize variational quantum circuits by jointly adapting their structure and parameters. Unlike fixed-structure approaches, QCE continuously evolves circuit designs to better solve optimization problems through a genetic-inspired algorithm. The evolutionary routine begins with a randomly initialized minimal circuit. At each iteration (generation), multiple variations (offspring) of the parent circuit are generated using mutation operations such as gate insertion, deletion, swapping, or parameter modification. Each offspring circuit is evaluated by measuring its performance against a predefined cost function, typically the expectation value of a Hamiltonian. The best-performing circuit is selected as the parent for the subsequent generation, ensuring progressive optimization, see Algorithm \ref{alg:qce}. This evolutionary mechanism not only bypasses gradient computations but also mitigates common issues in quantum circuit optimization such as barren plateaus and local minima, thereby enhancing robustness and solution quality.
\color{black}
\begin{algorithm}
\caption{Quantum Circuit Evolution (QCE)~\cite{QCEPROC} }
\label{alg:qce}
\begin{algorithmic}[1]
\STATE \textbf{Initialize:} Initial quantum circuit (parent) with a single random quantum gate [$R_x, R_y, R_z, CR_x, CR_y, CR_z, R_{xx}, R_{yy}, R_{zz}$]
\STATE \textbf{Copies:} Generate 4 offspring circuits
\STATE \textbf{Mutation:} Apply mutation with its respective probability:
\hspace{0.5cm} (insert (0.25), delete (0.25), swap (0.25), modify (0.25))
\STATE \textbf{Evaluate:} Evaluate each offspring by computing cost function
\STATE \textbf{Select:} Select the offspring with the lowest cost as the parent for the next generation
\STATE Repeat the evolution process until convergence or stopping criteria are satisfied
\end{algorithmic}
\end{algorithm}

\section{Introduction of the Adaptive Cost Function}
\begin{figure}
    \centering
   \includegraphics[width=0.47\textwidth]{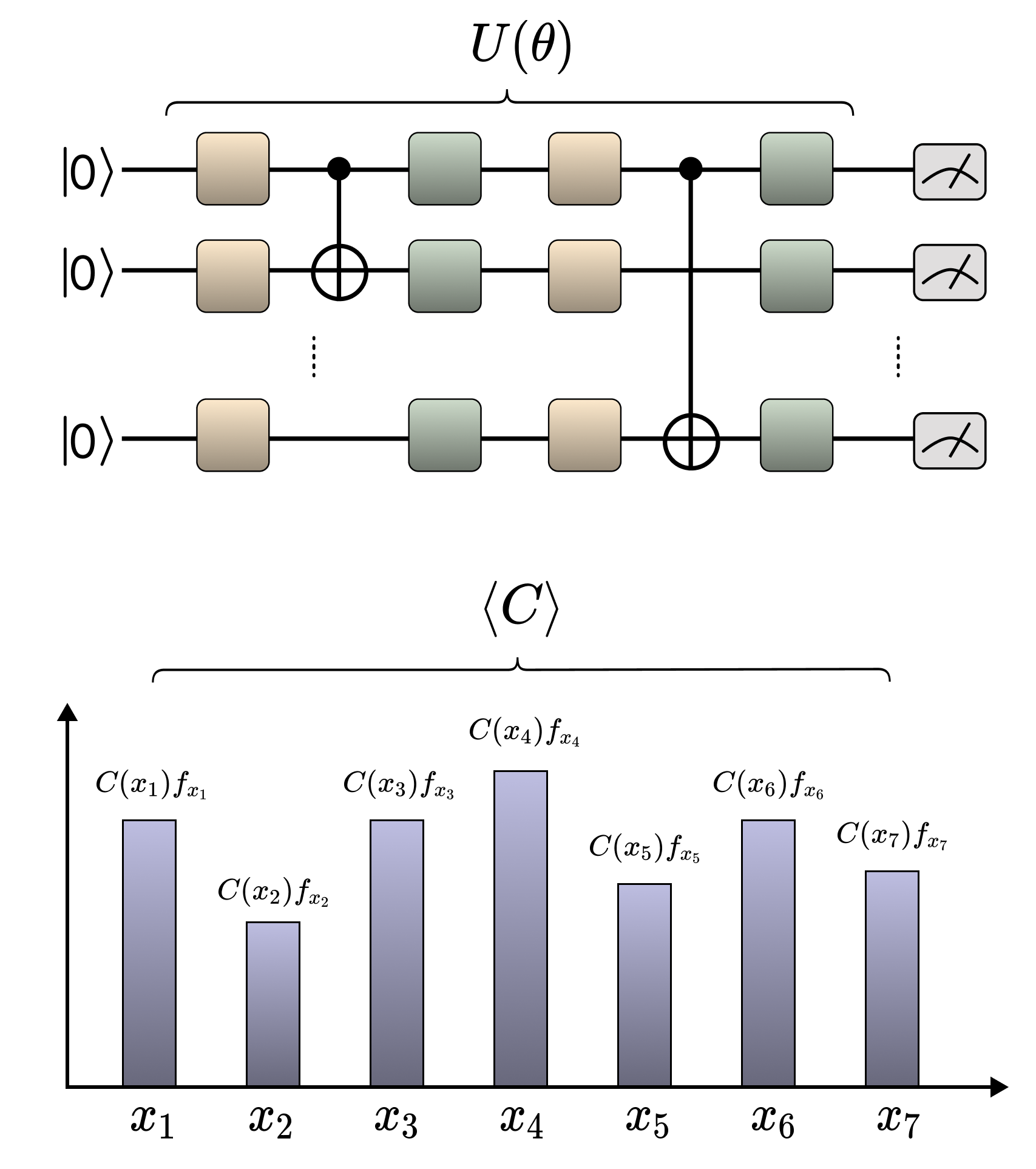}
    \caption{Illustration of a measurement performed on a generic circuit. The important factor is that each measured string can be associated with a probability associated with a value of the cost function in this string.This makes it clear how often each evaluation of the cost function receives per circuit (generation). Consequently, it is easy to see the influence of strings that are violations in the calculation of $\langle C\rangle$ described in eq. \ref{average}.}
   \label{fig:1}
\end{figure}
In this section we present the new approach used to accelerate the QCE convergence. The main feature of the proposed approach relies on modifications in the expectation value cost function considering what we call here violations of the penalties (constraints) of the QUBO. The conventional QCE uses a default cost function (DCF) as defined in \cite{QCEPROC} which is given by
\begin{equation}
 \langle {\cal H}_C\rangle=\langle \mathbf{0}|\mathbf{U}^{\dagger}{{\cal H}_{C}}\mathbf{U}|\mathbf{0}\rangle,
\label{DCF}
\end{equation}
where $\mathbf{U}$ is the circuit (which varies accordingly to the Algorithm \ref{alg:qce}) and ${\cal H}_C$ is the Ising Hamiltonian associated to eq. \ref{cost}.

We define the cost function in terms of an average of measures of circuit combined with eq. \ref{cost} without constructing the Ising Hamiltonian explicitly. Such a cost function is defined as (see Fig. \ref{fig:1})
\begin{equation}
    \langle C\rangle=\frac{1}{N_{t}}\sum_{\mathbf{x}}f_{\mathbf{x}}C(\mathbf{x}),
    \label{average}
\end{equation}
with $f_\mathbf{x}$ representing the measured frequency of the string (state) $\mathbf{x}$ and $N_{t}$ the total number of shots. The eq. \ref{average} is not a novelty properly since the QAOA method may use this approach with $f_{\mathbf{x}}(\theta)$, $\theta=(\beta,\gamma)$, which are parameters to be determined by classical optimization \cite{faihi2014qaoa}. Also, no classical optimization is demanded in the QCE. The main difference here relies upon  the QCE parameter independence because it is a variable circuit method, i.e., in our procedure the $f_{\mathbf{x}}(\mathbf{U})$ is circuit-dependent. At first glance eqs. \ref{average} and \ref{DCF} have a similar structure and, in fact, they have exactly the same performances. However, eq. \ref{average} presents a flexible structure that allows improving the performance of QCE with respect to avoiding convergence stagnation.
\subsection{Adaptive Cost Function Framework}
\begin{figure}
    \centering
   \includegraphics[width=0.5\textwidth]{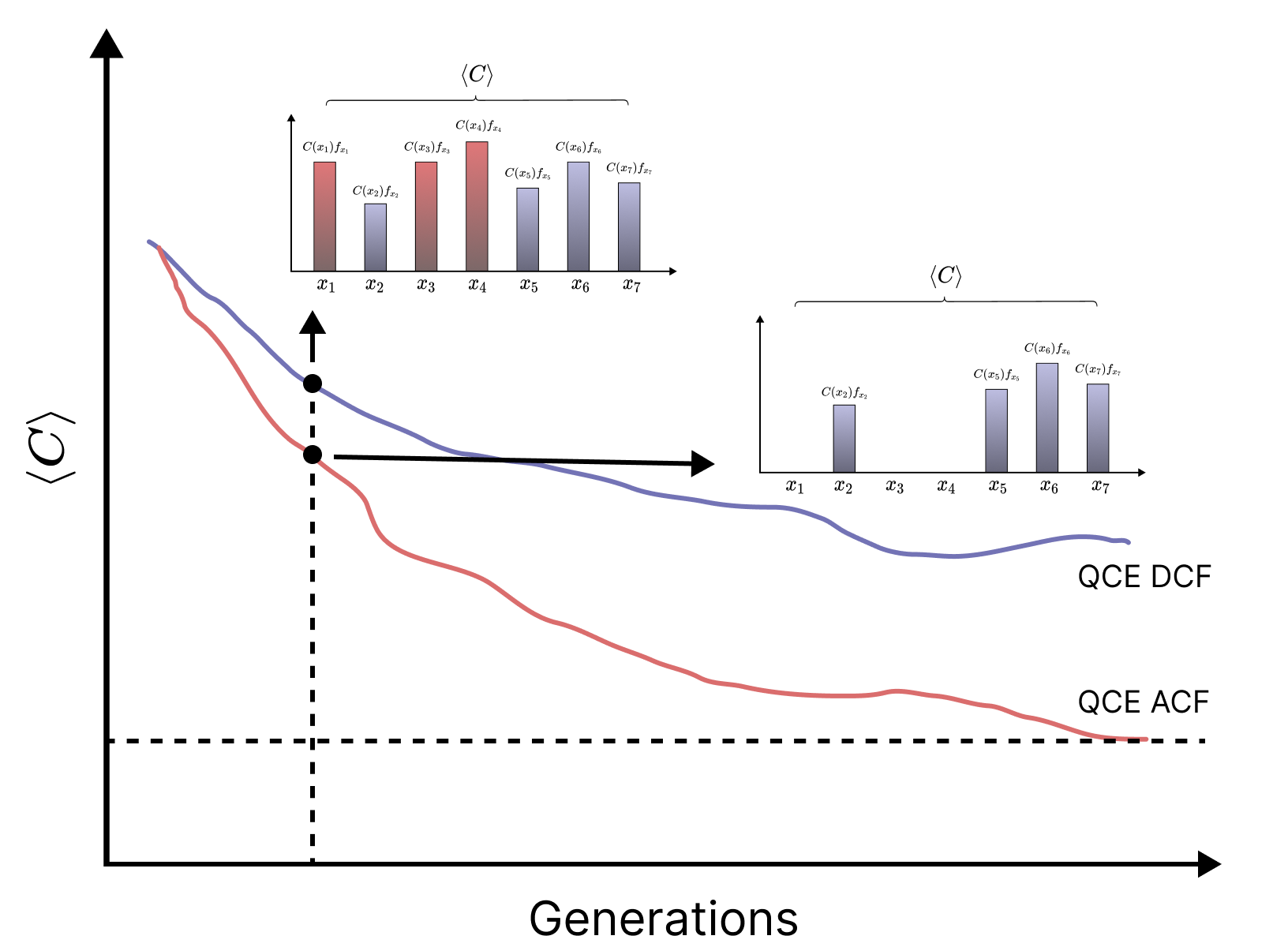}
    \caption{Illustration of DCF versus ACF behavior. The red boxes in the histogram represent constraint violations while the gray ones represent feasible strings. The ACF method discards violations that contribute to the stagnation of the cost function thus contributing to a faster convergence of the method.}
   \label{fig:2}
\end{figure}
The definition of the cost function given by eq. \ref{average} shows that contributions from strings that do not correspond to feasible solutions influence the value of the cost function in a way that hinders the convergence process. Taking into account that we do not use an optimization process to update the circuit, it is possible to introduce adaptations in the customized function in order to eliminate strings that compromise the method's convergence process. Strings that do not satisfy the problem constraints will be called violations, and strings that represent feasible solutions will be called feasible strings.

For each generation $j$, the measurement performed on the circuits may create a set ${\cal F}$ associated with the frequency of each feasible string and another set ${\cal V}$ of frequencies of the strings that are violations. The quantity $|{\cal V}|$ is the cardinality. 

Let us  consider $\cal M$ as the set of measured strings in the generation $j$ which contains more than one violation string. If we have  ${\cal M} \cap{\cal F} \neq 0$, we use 
The  cost function, dependent on generation $j$, defined as
\begin{equation}
    \langle C_j\rangle_{{\cal F}}=\frac{1}{(N_{t}-|{\cal V}|)}\sum_{\mathbf{x}_j\in {\cal F}}f_{\mathbf{x}_j}C(\mathbf{x}_j).
    \label{ACF_1}
\end{equation}
This formula works well when the circuit generated by QCE has some feasible strings among its measured strings since violations will be discarded, as illustrated in Fig. \ref{fig:2}. However,
the major flaw in eq. \ref{ACF_1} occurs when the generation produces a circuit with violations only. As a consequence, the adaptive cost function will be ill-defined since a division by zero will happen.

In these cases, we will reduce the space by discarding the violation that appears most frequently. Therefore, if we have $\cal M=\cal V$ and in this case ${\cal V}=({\cal V}-{\cal V}_{m})\cup{\cal V}_{m}$ where the last represents a subset with a larger frequency string. In this case, the cost function is
\begin{equation}
    \langle C_j\rangle_{{\cal V}}=\frac{1}{(N_{t}-|{\cal V}_{m}|)}\sum_{\mathbf{x}_j\in {({\cal V}-{\cal V}_{m})}}f_{\mathbf{x}_j}C(\mathbf{x}_j).
    \label{ACF_2}
\end{equation}

Finally, given $\cal M$, and considering the above discussion, it is possible to define for each generation the adaptive cost function as follows:
 \[
        \langle C_j\rangle=
           \begin{cases}
            \langle C_j\rangle_{{\cal F}}, & \text{if}\quad {\cal M} \cap{\cal F} \neq 0 \\
            \langle C_j\rangle_{{\cal V}}, & \text{otherwise}.
           \end{cases}
           \]

It is convenient to mention that the set $\cal M$ needs to contain more than one string in the case where there are only violations of the constraints. In this case, the algorithm imposes a condition in order to avoid this restriction. The adaptive cost function (ACF) is introduced in the step $4:$ in the Algorithm~\ref{alg:qce} replacing the default cost function defined in the eq. \ref{DCF}.
\section{Simulator Results}
Computational simulations were conducted on the quantum simulator KUATOMU at SENAI CIMATEC's High-Performance Computing Center (HPC). The computing infrastructure consists of 192 CPU processing cores with 384 threads, 3TB RAM, and four NVIDIA V100 32GB GPU accelerator cards. All implementations utilized the open-source Qiskit library.

Experiments were performed to evaluate QAOA and QCE-ACF algorithms for solving instances of the partitioning problem. The problem instances used in these experiments were originally provided by Svensson et al. \cite{database} and structured by R. Cacao et al. \cite{VQE_SHORT}.

We divided in two parts: noise-free and noise-induced scenarios, both detailed in the subsections below. Each experimental scenario was executed seven times. The algorithm performance was assessed using two metrics: execution time (in seconds) and an efficiency metric denoted as ratio , which is based on the mean expected value of the cost function. The  metric ranges between 0 and 1, indicating the proximity to the global minimum—an  value of 1 implies that the global minimum was successfully reached. In all our experiments, both algorithms (QAOA and QCE-ACF) consistently achieved an  value of 1; thus, execution time was utilized as the deciding factor for comparative analysis.

In Figure \ref{fig:3}, were included the results for QCE-DCF to highlight the comparative analysis of different methods. The Ratio metric frequently remained below 1, indicating that the approach did not reach the expected global minimum in optimization. This demonstrates that QCE-DCF did not achieve satisfactory solutions compared to the other analyzed approaches, reinforcing the superior performance of our approach for solving the studied problems.

\subsection{Noise-Free experiment}

\begin{figure}
    \centering
   \includegraphics[width=0.45\textwidth]{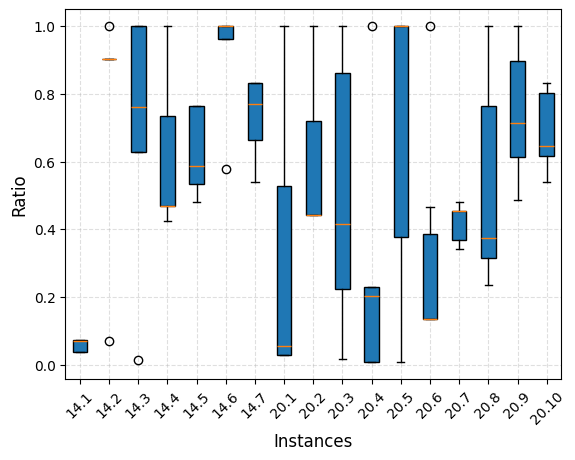}
    \caption{Results considering different instances of set partitioning problem solved using QCE-DCF. The average shows poor performance compared to the QAOA presented in the Table \ref{tab:res-noise-free}. We can notice some outliers in the instances $14.2$, $20.4$ and $20.6 $ that obtained convergence to the reference value.}
   \label{fig:3}
\end{figure}

For the noise-free experiments, simulations were conducted across 17 distinct instances of the partitioning problem, 7 instances of 14 qubits and 10 instances of 20 qubits considering $1024$ shots for the circuit measure. The QCE-DCF results are illustrated in Fig. \ref{fig:3} and the results for QAOA and QCE-ACF are summarized in Table \ref{tab:res-noise-free} illustrating the success ration and average execution time.

Let us begin by observing the originally proposed QCE-DCF which had its circuit mutation parameters adjusted according to Algorithm \ref{alg:qce} using the eq.~\ref{DCF}. The performance is illustrated in Fig. \ref{fig:3} and shows the method's difficulties in achieving regular and efficient performance. The averages show that the method presented convergence stagnation problems in several instances.

The QAOA implementation employed the standard model from the Qiskit Algorithms library, configured using the Cobyla optimizer (with details given in \cite{COBYLA}), one layer, and the Sampler estimator from Qiskit Primitives. The QCE-ACF algorithm was implemented as proposed and described in the preceding sections, utilizing the AerSimulator estimator provided by Qiskit Aer. The estimators were carefully chosen from the various available options within the Qiskit library to minimize execution times for both algorithms.

\begin{table}[htbp]
\caption{Noise-Free results}
\begin{center}
\begin{tabular}{|c|c|c|c|c|c|c|}
\hline
\textbf{Instances} & \multicolumn{2}{|c|}{\textbf{QAOA}}  & \multicolumn{2}{|c|}{\textbf{\rule{0pt}{10pt} QCE ACF}} \\
\cline{2-5} 
 & \myVSpace[1pt] \textbf{Ratio} & \textbf{Time (s)} 
  & \textbf{Ratio} & \textbf{Time (s)} \\
\hline
\myVSpace[2pt] 14.1  & 1.0 & 19.37   & 1.0 & 1.08 \\
14.2  & 1.0 & 18.03   & 1.0 & 0.63 \\
14.3  & 1.0 & 18.76   & 1.0 & 0.48 \\
14.4  & 1.0 & 17.50   & 1.0 & 0.95 \\
14.5  & 1.0 & 17.33   & 1.0 & 1.02 \\
14.6  & 1.0 & 19.65    & 1.0 & 1.09 \\
14.7  & 1.0 & 19.34   & 1.0 & 0.36 \\
\myVSpace[2pt] 20.1  & 1.0 & 1694.84  & 1.0 & 9.50 \\
20.2  & 1.0 & 1672.55  & 1.0 & 6.13 \\
20.3  & 1.0 & 1650.96  & 1.0 & 10.75 \\
20.4  & 1.0 & 1602.44  & 1.0 & 5.25 \\
20.5  & 1.0 & 1773.06  & 1.0 & 7.52 \\
20.6  & 1.0 & 1718.82  & 1.0 & 2.84 \\
20.7  & 1.0 & 1732.99  & 1.0 & 15.86 \\
20.8  & 1.0 & 1671.32  & 1.0 & 9.24 \\
20.9  & 1.0 & 1594.94  & 1.0 & 5.33 \\
20.10 & 1.0 & 1643.58  & 1.0 & 2.34 \\
\hline
\end{tabular}
\label{tab:res-noise-free}
\end{center}
\end{table}

The results for QAOA and QCE-ACF demonstrate a stark contrast in terms of execution time while maintaining similar solution quality. QAOA consistently achieves a Ratio of 1.0 across all instances, indicating that it successfully reaches the optimal or near-optimal solution. However, its execution times are significantly higher, especially for larger instances, where it reaches values exceeding 1,600 seconds. On the other hand, QCE-ACF also achieves a Ratio of 1.0 in all cases but does so with drastically lower execution times, often within a few seconds for smaller instances and remaining significantly below the time required by QAOA even for larger problem sizes. This suggests that QCE-ACF is a much more efficient approach in terms of computational cost while still maintaining high solution quality, making it a more suitable choice for large-scale optimization problems.

\subsection{Noise-induced experiment}

In the noise-induced scenario, due to computational constraints, simulations were limited to the seven 14-qubit problem instances (also considering $1024$ shots with  results presented in Table \ref{tab:res-noise}. The experimental setup was consistent with the noise-free experiments except for the introduction of noise.

Noise configurations were specifically chosen to assess algorithmic performance under practical perturbations. Errors were deliberately introduced to both the "cx" gates and measurement operations. The simulated quantum hardware was assumed to have ideal coupling, with all qubits interconnected, and noise effects were uniformly applied across all qubits. For QAOA, the AerSampler estimator from Qiskit Aer Primitives replaced the noise-free estimator to incorporate noise effects into the simulation. QCE-ACF maintained the use of the AerSimulator estimator from Qiskit Aer under noisy conditions.

\begin{table}[htbp]
\caption{Noisy results}
\begin{center}
\begin{tabular}{|c|c|c|c|c|}
\hline
\textbf{Instances} & \multicolumn{2}{|c|}{\textbf{ \rule{0pt}{10pt}QAOA}} & \multicolumn{2}{|c|}{\textbf{QCE ACF}} \\
\cline{2-5} 
 & \myVSpace[3pt] \textbf{Ratio} & \textbf{Time (s)} 
 & \textbf{Ratio} & \textbf{ Time (s)} \\
\hline
\myVSpace[2pt] 14.1   & 1.0 & 117.17  & 1.0 & 1.35  \\
14.2   & 1.0 & 134.45  & 1.0 & 3.21  \\
14.3   & 1.0 & 437.42  & 1.0 & 5.12  \\
14.4   & 1.0 & 201.25  & 1.0 & 6.90  \\
14.5   & 1.0 & 137.32  & 1.0 & 8.44  \\
14.6   & 1.0 & 268.78  & 1.0 & 28.96 \\
14.7   & 1.0 & 176.13  & 1.0 & 38.20 \\
\hline
\end{tabular}
\label{tab:res-noise}
\end{center}
\end{table}

Under noisy conditions, the performance differences between QAOA and QCE-ACF become even more pronounced. While both methods consistently achieve a Ratio of 1.0, indicating their ability to find optimal or near-optimal solutions, the execution times tell a different story. The QAOA algorithm experiences significantly higher computational costs, with runtimes ranging from 117.17 to 437.42 seconds, whereas QCE-ACF remains highly efficient, completing the same tasks in just 1.35 to 38.20 seconds. As the problem size increases, the disparity in execution time grows, with QAOA becoming increasingly expensive while QCE-ACF maintains a relatively stable performance. These findings suggest that QCE-ACF is not only computationally superior but also more robust in handling noise, making it a more viable approach for real-world quantum applications where hardware imperfections are inevitable.

\color{black}
\section{Comments and conclusions}
The QAOA method has been presented in the literature as a suitable tool for solving integer optimization problems.For the studied instances, it has some disadvantages in relation to our proposed approach. The first one is regarding circuit configuration, which is fixed and depends on the Ising Hamiltonian of the problem. The second relies on the classical optimization procedure to find the optimum circuit parameters. As shown in \cite{cerezop} such kind of circuits has an intrinsic difficulty to converge on a large scale even in shallow circuits. For the studied instances, considering the selected optimizer, QAOA obtained accurate results with execution time significantly increasing when the number of quits increases. 

The conventional QCE-DCF method achieves convergence to an answer for certain runs in several instances. However, for many other runs, the method would stagnate and fail to converge. One way to try to escape this is by introducing a large number of generations or more random mutations in the circuit. As a consequence,  in this case, the execution time for a single instance of $20$ qubits could exceed an hour without guaranteeing convergence for all runs. For this reason, its performance on average is poor compared to QAOA.

\begin{figure}[h!]
    \centering
   \includegraphics[width=0.45\textwidth]{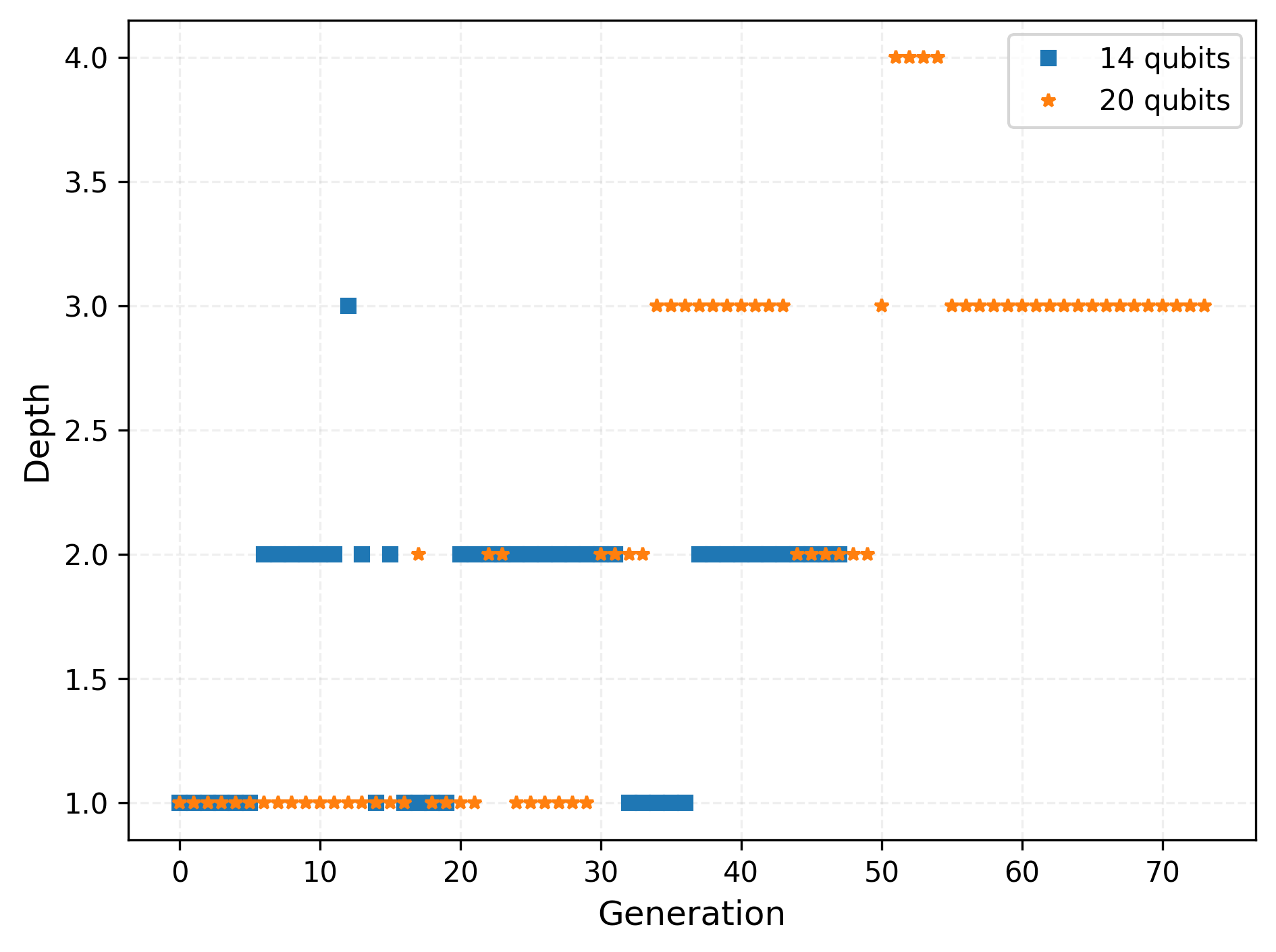}
    \caption{Evolution of circuit depth per generation. The QCE-ACF methodology allows faster convergence combined with shallower depth circuits. The depth used in this case is much more efficient than that of QAOA. For comparison purposes, in the case of $14$ qubits, QAOA required a circuit with depth $211$.}
   \label{fig:4}
\end{figure}

The introduction of the QCE-ACF framework completely changes this scenario. This approach obtained performances, in relation to execution time, superior to QAOA in both scenarios considering noise and without noise. Since the depth of generated circuits per generation remained shallow in most instances, there seems to be a tendency for circuit complexity to not scale exponentially with increasing qubits. In Fig. \ref{fig:4} it is possible to see the evolution of the depth of the circuit generated for each generation in two selected instances. 

Regarding verifications of violations and feasible strings after the measures, if the problem has $K$ constraints and $N$ variables (qubits), for $K$ independent of $N$, then checking for violations has $O(KN^2)$ operations with  polynomial order complexity demonstrating usefulness in larger problems. 

To conclude, the QCE-ACF method showed good convergence capabilities with fast execution time in the context of the studied instances, presenting encouraging results. The method also appears to simplify circuit evolution by obtaining efficient low-complexity configurations, looking quite promising for scale. However, such a scenario needs to be explored in detail in future work for larger instances, different problems, and other challenging benchmarks. 
\section{Acknowledgments}
This work has been partially supported by QuIIN - EMBRAPII CIMATEC Competence Center in Quantum Technologies, with financial resources from the PPI IoT/Manufatura 4.0 of the MCTI grant number 053/2023, signed with EMBRAPII. Also, this work was partially financed by CNPq (Grant Numbers: 305096/2022-2, Marcelo A. Moret). We also acknowledge the use of the Qiskit software development kit, developed by IBM, which was instrumental in the implementation and simulation of quantum circuits in this research.
\textbf{Code availability statement:} The code used in this work is available at: \href{https://github.com/brunooziel/quantum-circuit-evolution-with-an-adaptive-cost-function}{https://github.com/brunooziel/quantum-circuit-evolution-with-an-adaptive-cost-function}.

\bibliographystyle{IEEEtran}
\bibliography{references}
\end{document}